\documentclass[aps,prc,preprint,groupedaddress,showpacs,preprintnumbers,
amsmath,amssymb]{revtex4}
\usepackage{graphicx}% Include figure files
\usepackage{dcolumn}% Align table columns on decimal point
\usepackage{bm}% bold math

\begin{document}

% Use the \preprint command to place your local institutional report
% number in the upper righthand corner of the title page in preprint mode.
% Multiple \preprint commands are allowed.
% Use the 'preprintnumbers' class option to override journal defaults
% to display numbers if necessary
%\preprint{}

%Title of paper
\title{The  
$^4$He$(e,e^\prime p)^3$H Reaction with Full Final--State Interaction}

% repeat the \author .. \affiliation  etc. as needed
% \email, \thanks, \homepage, \altaffiliation all apply to the current
% author. Explanatory text should go in the []'s, actual e-mail
% address or url should go in the {}'s for \email and \homepage.
% Please use the appropriate macro foreach each type of information

% \affiliation command applies to all authors since the last
% \affiliation command. The \affiliation command should follow the
% other information
% \affiliation can be followed by \email, \homepage, \thanks as well.
\author{Sofia Quaglioni$^1$, Victor D. Efros$^2$, Winfried Leidemann$^1$ 
and Giuseppina Orlandini$^1$}
%\email[]{Your e-mail address}
%\homepage[]{Your web page}
%\thanks{}
%\altaffiliation{}
\affiliation{$^1$Dipartimento di Fisica, Universit\`a di Trento, and 
Istituto
Nazionale di Fisica Nucleare, Gruppo Collegato di Trento, I-38050 Povo, 
Italy\\
$^2$Russian Research Centre ``Kurchatov Institute'', 
Kurchatov 
Square, 1, 123182 Moscow, Russia}

%Collaboration name if desired (requires use of superscriptaddress
%option in \documentclass). \noaffiliation is required (may also be
%used with the \author command).
%\collaboration can be followed by \email, \homepage, \thanks as well.
%\collaboration{}
%\noaffiliation

\date{\today}

\begin{abstract}
% insert abstract here
An {\it ab initio} calculation of the $^4$He$(e,e^\prime p)^3$H 
longitudinal response is presented. The use of the integral transform
method with a Lorentz kernel has allowed
to take into account the full four--body final state interaction (FSI). 
The semirealistic nucleon-nucleon potential MTI--III and the Coulomb force are
the only ingredients of the calculation. 
The reliability of the direct knock--out hypothesis 
is discussed both in parallel and in non parallel kinematics.  
In the former case it is found that lower missing 
momenta and higher momentum transfers are preferable to minimize effects
beyond the plane wave impulse approximation (PWIA). 
%[??? Contrary to previous estimates based on approximate approaches,
%we find that FSI enhances the PWIA results at higher missing momenta, raising some 
%questions about the extractions of momentum distributions and 
%spectroscopic factors ???]. 
Also for non parallel kinematics the role of antisymmetrization and final state interaction
become very important with increasing missing momentum, raising  
doubts about the possibility of extracting momentum distributions and 
spectroscopic factors. 
The comparison with experimental results in parallel kinematics,
where the Rosenbluth separation has been possible, is discussed.

\end{abstract}

% insert suggested PACS numbers in braces on next line
% insert suggested keywords - APS authors don't need to do this
%\keywords{}

%\maketitle must follow title, authors, abstract, \pacs, and \keywords

\maketitle

% body of paper here - Use proper section commands
% References should be done using the~\cite, \ref, and \label commands
%\section{}
% Put \label in argument of \section for cross-referencing
%\section{\label{}}
%\subsection{}
%\subsubsection{}

\section{Introduction}

Numerous experimental as well as theoretical investigations of $(e,e^\prime p)$ 
exclusive reactions, in both light and heavy nuclei, have been performed
extensively in the past, with the aim to extract information about  
the structure of these systems 
\cite{FM:1984,LBB:1986,SBB:1986,SBB:1987,UBB:1987,BBC:1987,RBB:1988,MBD:1988,BBE:1988,SBJ:1988,SBJ:1988b,HBJ:1988,LBB:1989,MBB:1989,LBJ:1990,BBB:1991,DBB:1993,LCH:1994,GBB:1994,LBB:1998} .
In particular one has tried to access ground state properties of the target 
nucleus like spectroscopic factors, shell momentum distributions etc.
However, it is well known that such quantities can only be obtained under 
the hypothesis that the reaction mechanism 
is dominated by a direct knock--out of the proton and neglecting the interaction
in the final state. Such assumptions are usually
considered more and more plausible as the momentum transferred by the electron 
to the system increases and allows to probe ``single nucleon'' physics.
In many nuclei the experimental one--body knock--out spectra 
indeed show very nicely pronounced peaks, hinting to an 
independent motion of the nucleons in such systems. 
In these cases shell momentum distributions and spectroscopic factors
have been extracted trying to estimate FSI effects in various ways.
Spectroscopic factors which are found smaller than 1 (and they are often 
found considerably far from 1) are interpreted as due to large 
``correlation effects'' induced by the residual interaction in the ground state of
the system.

Unfortunately up to recently the two  fundamental assumptions mentioned 
above could not be checked, because of the 
impossibility to solve the man--body problem in a quantum mechanical consistent way
both for ground and continuum states. 
The recent progress made in few--body physics allows us to start
investigating these assumptions. For $A=2$ and 3 the calculations 
are fully under control ~\cite{ALT:2005,GWS:2004} both for ground and continuum states,
and the problem has been investigated. However, the features of such systems 
are often considered too different from those of a typical ``many--body'' nucleus, 
to be taken as testing grounds for validating assumptions on heavier systems. 

In the last decade it has been demonstrated that the procedure to calculate reactions
with the help of integral
transforms originally proposed in 1985~\cite{EFROS:1985}
can be successfully applied 
in order to overcome the longstanding stumbling block which prevented
{\it ab initio} calculations of high energy reactions involving four 
nucleons and more~\cite{ELO:1997A,ELO:1997B,BELO:2001,BMBLO:2002,BABLO:2004}.
This has been possible thanks to the integral transform with the Lorentz kernel
(we will denote it by LIT)
proposed in Ref.~\cite{ELO:1994} and recently applied also to the two--body 
break--up of the four--body 
system in Ref.~\cite{QBELO:2004}.
Thus it is possible now to treat the full dynamics of a reaction to continuum 
in a nucleus
whose features (binding energy and density) are certainly much closer to 
those of heavier systems than deuteron and triton or $^3$He.

It is the purpose of this work to perform a model study of the role of the 
full treatment of the 
interaction in the four--body dynamics in the $^4$He$(e,e^\prime p)^3$H reaction
and to discuss the plausibility of the direct knock--out and plane wave 
assumptions. To this aim we use the semirealistic potential MTI--III \cite{MT:1969}
and concentrate on the longitudinal response 
function, where meson exchange currents effects are negligible.
This response is accessible experimentally if one performs a Rosenbluth 
separation  in parallel kinematics, and has been measured in a number
of experiments ~\cite{DBB:1993,MBB:1989}. Though the use of a semirealistic potential
does not allow to draw precise conclusions we believe that a comparison with data 
may be instructive and we will also comment on that.

In Sec.~II the expression for the $(e,e^\prime p)$ cross section is recalled
and the formalism describing the integral transform approach with a Lorentz 
kernel to exclusive reactions is
reviewed. Results are given in Sec.~III while conclusions are summarized in Sec.~IV.

\section{general formalism}

\subsection{Cross Section}

The sixfold electrodisintegration cross section of $^4$He
into the two fragments $p$ and $^3$H 
is given by~\cite{Don:1985,BGP:1993} 
\begin{eqnarray}
\frac{d^6\sigma_{p,t}}{dE^\prime d\Omega_{e^\prime} d{\bf p}_p} &=& \sigma_M \left[ 
V_L S_L(\omega,q,p_p,\theta_p) + V_T S_T(\omega,q,p_p,\theta_p)+ 
V_{LT} S_{LT}(\omega,q,p_p,\theta_p) \cos\phi\right.\nonumber\\
&&\left. 
+ V_{TT} S_{TT}(\omega,q,p_p,\theta_p) \cos 2\phi\right]\,.
\label{sigma}
\end{eqnarray}
Here $E^\prime$ and $\Omega_{e^\prime}$ denote energy and solid angle of the 
electron
after the reaction, $\sigma_M$ is the Mott cross section and $\phi$ denotes the 
angle between
the electron and ejectile planes. Energy and momentum 
transferred from the electron to the 
nuclear system are denoted as $\omega$
and ${\bf q}=q\hat{\bf q}$. The quantity 
${\bf p}_p=(p_p,\Omega_p)$ denotes the momentum of the proton 
detected in coincidence with the electron, and $\theta_p$ is the angle
between the outgoing proton and $\hat{\bf q}$.  
The $V_{\beta}$ are kinematical 
coefficients and the nuclear dynamics is contained by the 
structure functions $S_{\beta}$. 
%The latter include the energy conservation delta function (which represents as 
%well a relation among the  arguments of the structure functions). 

Integration over $p_p$ 
leads to the fivefold cross section
%%%%%%%%%%%%%%%%%%%%%%%%%%%%%%%%%%%%%%%%%%%%%%%%%%%%%%%%%%%%%%%%%%%%%%%%%%%%%%%%%%%%%
%Here $E^\prime$ and $\Omega_{e^\prime}$ denote energy and solid angle of the electron
%after the reaction, ${\bf p}_p=(p_p,\Omega_p)$ is the momentum of the proton 
%detected in coincidence with the electron, $\sigma_M$ is the Mott cross section 
%and $\phi$ denotes the angle between
%the electron and ejectile planes. The $V_{\alpha}$ are kinematical 
%coefficients and the nuclear dynamics is contained in the 
%structure functions $S_{\alpha}$. They are functions of the energy $\omega$
%and momentum ${\bf q}=q\hat{\bf q}$ transferred from the electron to the nuclear 
%system as well as of the modulus $p_p$ and angle $\theta_p$ (with respect to 
%$\hat{\bf q}$) of the outgoing proton.
%Integration over $p_p$ leads to the fivefold cross section
%%%%%%%%%%%%%%%%%%%%%%%%%%%%%%%%%%%%%%%%%%%%%%%%%%%%%%%%%%%%%%%%%%%%%%%%%%%%%%%%%%%%%%
\begin{eqnarray}
\frac{d^5\sigma_{p,t}}{dE^\prime d\Omega_{e^\prime} d\Omega_p} &=&\int{\frac{d^6\sigma_{p,t}}
{dE^\prime d\Omega_{e^\prime} d{\bf p}_p} \frac{p_p^2}{\left|\frac{\partial E_m}
{\partial p_p}\right|}dE_m}\nonumber\\
&=&\sigma_M  \frac{p_p^2}{\left|\frac{\partial E_m}{\partial p_p}\right|} \left[ 
V_L F_L(\omega,q,\theta_p) + V_T F_T(\omega,q,\theta_p)\right.\nonumber\\
&&\left. + V_{LT} F_{LT}(\omega,q,\theta_p) \cos\phi+ V_{TT} F_{TT}(\omega,q,\theta_p) 
\cos 2\phi)\right]\,,
\end{eqnarray}
where $E_m=\omega-T_p-T_t$ represents the missing energy ($T_p$ and $T_t$ being the 
proton and triton kinetic energies).  
%\mbox{$\delta(E_m-E_m^0)$}
The new structure functions $F_\beta$ are simply given by
\begin{equation}
F_\beta(\omega,q,\theta_p)=\int{S_\beta(\omega,q,p_p,\theta_p) dE_m}\,.
\end{equation}
Notice that, since $S_\beta(\omega,q,p_p,\theta_p)$ include the energy conserving 
$\delta$--function, the integration over $E_m$ fixes a unique value of $p_p$ for
each combination of $\omega,q$ and $\theta_p$.
 
The total contribution of the $p,t$ disintegration channel to the inclusive
 cross section is
\begin{equation}
\frac{d^3\sigma_{p,t}}{dE^\prime d\Omega_{e^\prime}}=\sigma_M \left[ 
V_L R^{p,t}_L(\omega,q) + V_T R^{p,t}_T(\omega,q)\right]\,,
\end{equation}
with
\begin{equation}\label{RLpt}
R^{p,t}_\beta(\omega,q)=\int{\frac{2\pi p_p^2}{\left|\frac{\partial E_m}
{\partial p_p}\right|} F_\beta(\omega,q,\theta_p) \sin\theta_p d\theta_p}\,.
\end{equation}

In what follows we concentrate on the 
longitudinal response $F_L(q, \omega,\theta_p)$, representing the response 
of the system to the electron-nuclear charge interaction. This  
can be written as  
\begin{equation}\label{FL}
{F_L(q, \omega,\theta_p)=\sum _{M_t,M_p}
\left|\left\langle\Psi_{p,t}^-(E_{p,t})
\left|\hat \rho({\bf q})\right|\Psi_{\alpha}\right\rangle\right|^2 }\,,
\end{equation}
where the four--body ground state  
is denoted by $\Psi_{\alpha}$, and
$\Psi^-_{p,t}$ is the continuum final--state 
of the minus type  pertaining to the proton--triton channel~\cite{GW:1964}
with the relative proton--triton momentum ${\bf k}=k\hat{\bf k}$. The sum goes over 
the projections $M_t$ and $M_{\mathit p}$ of the 
fragment total angular momenta in the final state. 
The continuum states $\Psi^{-}_{{\mathit p},t}$ are normalized 
to $\delta({\bf k}-{\bf k}^\prime)\delta_{M_tM^{\prime}_t}
\delta_{M_{p}M^{\prime}_{p}}$.
The quantity $E_{p,t}$ is the final state intrinsic energy
\begin{equation}\label{Ept}
E_{p,t}=\frac{k^2}{2\mu} + E_t\,,
\end{equation}
where $\mu$ is the reduced mass of the proton-triton system and $E_t$ denotes 
the $^3$H ground state energy. 

%$E_{p,t}$  is
%the input of the non relativistic calculation of the matrix element. 
%In comparing results with data one can determine it from the experimental
%value of $\omega$ from the relation 
%\begin{equation}\label{Ept1}
%E_{p,t}=\omega-\frac{q^2}{2 (m_p+m_t)}+E_{\alpha}\,,
%\end{equation}
%where $m_p$ and $m_t$ are the proton and triton masses, respectively. 
%Alternatively one can determine $k$ from the experimental values of ${\bf p}_p$
%and $q$, via the relative momentum definition
%\begin{equation}\label{k}
%{\bf k}= \mu ({{\bf p}_p\over m_p} - {({\bf q}-{\bf p}_p)^2\over m_t})\,,
%\end{equation}
%and obtain $E_{p,t}$ from Eq.~(\ref{Ept}).
%Since the latter is a non relativistic relation (consistent with the 
%Schr\"dinger Hamiltonian)  the two different procedures 
%generally lead to two different values of $E_pt$. 
%The results we will present in section III
%will make use of relation (\ref{k}), preferring momentum to energy conservation 
%(as it is usually done, e.g. in the determination
%of NN potentials). 

The initial and final states are connected by the nuclear charge operator 
$\hat \rho$ which we take in its non relativistic form 
%As in~\cite{ELO97,BELO01} only  transitions induced by the unretarded 
%dipole 
%operator,
%\begin{equation}
%D_z =\sum_{j=1}^{4}{\frac{1+\tau_{j}^{3}}{2}z_{j}} \,
%\label{dipole}
%\end{equation}
%are taken into account. 
\begin{equation}\label{rho}
\hat \rho({\bf q}) =\sum_{j=1}^{4}{G_E^p\,
\frac{1+\tau_{j}^{3}}{2}}
\exp{(i {\bf q}\cdot {\bf r}_j}) \,.
\end{equation}
Here $\tau^3_j$ denotes the third component 
of the $j$-th nucleon isospin, ${\bf r}_j$ represents the position of
the $j$-th nucleon with respect to the center of mass of the four--body
system and $G_E^p$ is the proton electric form factor. In comparing our results 
with experimental data we will use the proton form factor 
$\tilde G_E^p=G_E^p/(1+(q^2-\omega^2)/4 m_p^2)^{1/2}$ 
(containing first order relativistic correction) with
$G_E^p$ in the usual dipole parametrization.

The main difficulty in the calculation of $F_L$ is 
represented
by the  continuum  wave function $\Psi^-_{p,t}
(E_{p,t})$ in the transition matrix element 
\begin{equation}\label{Tpt}
T_{p,t}(E_{p,t})=
\left\langle\Psi_{p,t}^-(E_{p,t})\left|\hat \rho\right|
\Psi_{\alpha}\right\rangle~.
\end{equation}
With the integral transform method \cite{EFROS:1985} with the Lorentz kernel 
\cite{ELO:1994, LL:2000} 
one is able to perform an {\it ab initio} calculation of this   
transition matrix element in a large energy range 
without dealing with the continuum solutions of the four--body 
Schr\"odinger equation. How this is possible has been described in 
Ref.~\cite{EFROS:1985} 
and will be briefly summarized
in the next subsection. Further details can be found 
in~\cite{EFROS:1993,EFROS:1999,LL:2000,QBELO:2004}.

\subsection{The LIT Method for Exclusive Reactions}

The LIT approach to exclusive reactions consists in calculating 
transition 
matrix element of the perturbation $\widehat{O}$ between the initial 
($\Psi_0$) and final ($\Psi^-_f$) states
\begin{equation}\label{TfEf}
T_{f}(E_f)=\left\langle\Psi^-_f(E_f)\left|{\widehat O}\right|
\Psi_0\right\rangle~,
\end{equation} 
without calculating $\Psi^-_f(E_f)$.
%The calculation of such a matrix element can be carried out with the 
%LIT method as outlined in the following~\cite{Ef85,LL00}.

In general denoting  with $a$ and $b$ the two fragments containing $n_a$ 
and  $n_b=A-n_a$ nucleons, respectively and with $H$ the full nuclear 
Hamiltonian, we have the following formal 
expression for $\Psi^-_{f=a,b}(E_{f=a,b})$ in terms of the ``channel state'' 
$\phi_{f=a,b}^-(E_{f=a,b})$~\cite{GW:1964}
\begin{equation}\label{Psiminus}
\left|\Psi^-_{a,b}(E_{a,b})\right\rangle={\widehat{\mathcal A}}
\left|\phi_{a,b}^-(E_{a,b})\right\rangle+\frac{1}{E_{a,b}-i\varepsilon-H}
\widehat{\mathcal A}{\mathcal V}\left|\phi_{a,b}^-(E_{a,b})\right\rangle\,,
\end{equation}
where $\widehat{\mathcal A}$ is an antisymmetrization operator.
In case that at least one of the fragments is chargeless the channel wave function
 $\phi_{a,b}^-(E_{a,b})$ is the product of the internal wave 
functions of the fragments 
and of their relative free motion. Correspondingly, ${\mathcal V}$ in 
Eq.~(\ref{Psiminus}) is the sum of all interactions between particles 
belonging to different fragments.  If both fragments are charged, like in our case, 
$\phi_{a,b}^-(E_{a,b})$ is chosen to account for the average Coulomb 
interaction between them, and the plane wave describing their
relative motion is replaced by the Coulomb function of the 
minus type. Correspondingly, ${\mathcal V}$ in 
Eq.~(\ref{Psiminus}) is the sum of all interactions between particles 
belonging to different fragments after subtraction of the average Coulomb 
interaction, already considered via the Coulomb function.
We write $\phi_{a,b}^-(E_{a,b})$ in the partial wave  
expansion form 
\begin{equation}\label{Phiminus}
\phi^-_{a,b}(E_{a,b})=\frac{\Phi_a(1, ...,n_a)\Phi_b(n_a+1, ...,A)}
{(2\pi)^{3/2}}~4\pi\sum_{\ell=0}^{\infty}
\sum_{m=-\ell}^{\ell}{i^{\ell}e^{-i\delta_{\ell}(k)}\frac{w_{\ell}
(k;r)}{kr}Y_{\ell m}(\Omega_r)Y^*_{\ell m}(\Omega_k)\,.}
\end{equation}
Here $\Phi_a(1, ..., n_a)$ and $\Phi_b(n_a+1, ..., A)$ are the internal 
wave functions of the fragments, 
${\mathbf r}=(r,\Omega_r)={\mathbf R}^a_{cm}
-{\mathbf R}^b_{cm}$ represents the distance between them, and the 
energy of the relative motion is $k^2/ 2\mu=E_{a,b}-E_a-E_b$, where $E_a$ 
and $E_b$ are the fragment ground state energies. The functions $w_{\ell}(k;r)$ 
are the regular Coulomb wave functions of order $\ell $, and 
$\delta_{\ell}(k)$ are the Coulomb phase shifts~\cite{GW:1964}.   
The internal wave functions of the fragments are assumed to be 
antisymmetrized and normalized to unity, so that the properly normalized 
continuum wave function in Eq.~(\ref{Psiminus}) is obtained via application 
of the antisymmetrization operator. For $n_a=1$ this has the form
\begin{equation}\label{A}
{\widehat{\mathcal A}}=\frac{1}{\sqrt{A}}\left[1-\sum_{j=2}^{A}{\mathcal P}_{1j}\right]~,
\end{equation}    
where ${\mathcal P}_{ij}$ are particle permutation operators~\cite{GW:1964}. 

When one inserts Eq.~(\ref{Psiminus}) into Eq.~(\ref{TfEf})  the transition 
matrix element becomes the sum of two pieces, a Born term,
\begin{equation}\label{TBorn}
T_{a,b}^{Born}(E_{a,b})=\left\langle\phi_{a,b}^-(E_{a,b})\left|
{\widehat{\mathcal A}}~{\widehat O}\right|\Psi_0\right\rangle~,
\end{equation}
and a  FSI dependent term, 
\begin{equation}\label{TFSI1}
T_{a,b}^{FSI}(E_{a,b})=\left\langle\phi_{a,b}^-(E_{a,b})\left|
{\mathcal V}{\widehat{\mathcal A}}\frac{1}{E_{a,b}+i\varepsilon-H} 
{\widehat O}\right|\Psi_0\right\rangle~.
\end{equation}
While the Born term is rather simple to deal with, the determination  
of the FSI dependent matrix element is rather complicated. Within the LIT 
approach this term is treated as outlined in the following. 

In Eq.~(\ref{TFSI1}) one inserts the completeness relation of the 
Hamiltonian eigenstates 
$\Psi_\nu(E_\nu)$ (labelled  by channel quantum 
numbers $\nu$ and normalized as $\langle\Psi_{\nu}|\Psi_{\nu^{\,\prime}}\rangle=
\delta(\nu-\nu^{\,\prime})$)
%Using the completeness relation of the set 
%$\Psi_\nu(E_\nu)$ the matrix element $T_{a,b}^{FSI}(E_{a,b})$ can be 
%written as: 
\begin{equation}\label{TFSI2}
T_{a,b}^{FSI}(E_{a,b})=\sum\!\!\!\!~\!\!\!\!\!\!\int d\nu\,
\langle\phi_{a,b}^-(E_{a,b})|{\mathcal V}
{\widehat{\mathcal A}}|\Psi_\nu(E_\nu)\rangle\frac{1}{E_{a,b}+i
\varepsilon-E_\nu}\langle\Psi_\nu(E_\nu)| 
{\widehat O}|\Psi_0\rangle\,.\\
\end{equation}
Defining $F_{a,b}(E)$ as
\begin{equation}\label{F}
F_{a,b}(E)=\sum\!\!\!\!~\!\!\!\!\!\!\int d\nu\left\langle\phi_{a,b}^-
(E_{a,b})\left|{\mathcal V}{\widehat{\mathcal A}}\right|\Psi_\nu(E_\nu)
\right\rangle\left\langle\Psi_\nu(E_\nu)\left|{\widehat O}\right|\Psi_0
\right\rangle\delta(E-E_\nu)~,
\end{equation}
one has
\begin{eqnarray}\label{TFSI3}
T_{a,b}^{FSI}(E_{a,b})&=&\int_{E_{th}^-}^{\infty}{\frac{F_{a,b}(E)}
{E_{a,b}+i\varepsilon-E} 
dE}~=~-i\pi F_{a,b}(E_{a,b})+{\mathcal P}\int_{E_{th}^-}^{\infty}
{\frac{F_{a,b}(E)}{E_{a,b}-E}dE}~,
\end{eqnarray}
where $E_{th}$ is the lowest excitation energy in the system i.e. the 
break--up threshold energy. 

The function $F_{a,b}$ contains information on all the 
eigenstates $\Psi_\nu$ for the whole eigenvalue spectrum of $H$. It is
obtained by its Lorentz integral transform
\begin{equation}
L\left[F_{a,b}\right](\sigma)=
\int_{E_{th}^-}^{\infty}
{\frac{F_{a,b}(E)}
{(E-\sigma_R)^2+\sigma_I^2}~dE}~=~
\left\langle {\widetilde \Psi}_2(\sigma)
\left|\right.{\widetilde
\Psi}_1(\sigma)\right\rangle~,
\label{mod}
\end{equation}
where
\begin{equation}\label{Psitilda12}
{\widetilde\Psi}_1(\sigma)=(H-\sigma)^{-1}{\widehat O}\left|\Psi_0
\right\rangle,\qquad
{\widetilde\Psi}_2(\sigma)=(H-\sigma)^{-1}{\widehat {\mathcal A}}
{\mathcal V}
|\phi_{a,b}^-(E_{a,b})\rangle
\end{equation}
and $\sigma=\sigma_R+i\sigma_I$.
Equation (\ref{mod}) shows that $L\left[F_{a,b}\right](\sigma)$ can 
be calculated  without explicit knowledge of $F_{a,b}$, provided that 
one solves the two equations 
\begin{eqnarray}
(H-\sigma)\left|{\widetilde\Psi}_1\right\rangle&=&{\widehat O}
\left|\Psi_0\right\rangle\label{psitilde1}~,\\
(H-\sigma)\left|{\widetilde\Psi}_2\right\rangle&=&{\widehat 
{\mathcal A}}{\mathcal V}|\phi_{a,b}^-(E_{a,b})\rangle\,,
\label{psitilde2}
\end{eqnarray}
which differ in the source terms only. 
It is easy to show that ${\widetilde\Psi}_1$ and ${\widetilde\Psi}_2$ have 
finite norms. 
When solving Eqs.~(\ref{psitilde1}) and (\ref{psitilde2}) 
it is sufficient to require that the solutions are localized, and no 
other boundary conditions are to be imposed. Therefore ``bound state''
techniques can be applied. 

We use an expansion over 
a basis set of localized functions 
consisting of correlated hyperspherical harmonics
(CHH) multiplied by hyperradial functions. 
As discussed in~\cite{BELO:2001} for the case of the total $^4$He 
photoabsorption cross section, special attention has to be paid to the 
convergence of such expansions. A rather large number of basis states 
is necessary in order to reach convergence, thus leading to large 
Hamiltonian matrices. Instead of using a time consuming inversion method 
we directly evaluate the scalar products in (\ref{mod}) with the Lanczos 
technique as explained in Ref.~\cite{MBLO:2003}.

After having calculated $L[F_{a,b}](\sigma)$ one obtains the function 
$F_{a,b}(E)$, and thus $T_{a,b}(E_{a,b})$, via the inversion of the LIT, 
as described in~\cite{ELO:1999}.

In the next section results obtained by means of Eq.~(\ref{TBorn})
will be labelled by PWIAS. The label PWIA will indicate
that in Eq.~(\ref{TBorn}) the antisymmetrization operator ${\mathcal A}$
has been neglected.
We remind the reader that in this case the structure function $F_L^{p,t}$ turns out to be  
factorized in terms of the proton form factor and a function 
$n(|{\bf q}-{\bf p_p}|)$, which 
is the Fourier transform of the overlap between the $^4$He and $^3$H 
ground state wave functions.

\section{results}

As already mentioned, the ground states of $^4$He and $^3$He  as well as the 
$\tilde \Psi$ in Eqs.~(\ref{psitilde1}) and (\ref{psitilde2}) are calculated
 using the CHH expansion method. 
In order to speed up the convergence, state independent correlations are introduced 
as in~\cite{ELO:1997A}. We use the MTI--III
\cite{MT:1969} potential and identical CHH expansions for the ground state 
wave functions of $^4$He and of the three--nucleon systems as in~\cite{BELO:2001} 
and~\cite{ELO:1997C}, respectively.

We calculate the transition matrix elements~(\ref{TBorn}) and~(\ref{TFSI1}) in the 
form of partial wave expansions. When one substitutes the expansion (\ref{Phiminus}) 
and the expansion
\begin{equation}\label{rhoexp}
{\hat\rho}({\bf q})=\sum_{LM}Y_{LM}^*(\Omega_q){\hat\rho}_{LM}(q)
\end{equation}
of the charge operator (\ref{rho}) into the Born amplitude (\ref{TBorn}) and 
into the
right--hand sides of Eqs.~(\ref{psitilde1}) and (\ref{psitilde2})
one finds that in our case 
of central NN forces the transition matrix element (\ref{Tpt}) turns into
a sum over over $L$ ($L$ is equal to the $l$ in (\ref{Phiminus})) 
of partial transition matrix elements multiplied
by the factors 
\begin{equation}\label{legendre}
\sum_{M=-L}^LY_{LM}(\Omega_k)Y_{LM}^*(\Omega_q)=(4\pi)^{-1}(2L+1)
P_L({\hat {\bf k}}\cdot{\hat {\bf q}}).
\end{equation}
These factors determine the dependence of the cross section on $\theta_p$.
The dynamic equations are split with respect to orbital momentum $L$ and 
they are 
$M$--independent. The multipole transitions of the charge operator in 
Eq.~(\ref{rho}) are taken into account up to a maximal value of $L_{Born}=20$ 
and $L_{FSI}=6$ for the Born and FSI terms, respectively. (The relatively low value
of $L_{FSI}$ is due to the fact that FSI does not affect the final state 
 higher partial waves significantly.)
Correspondingly 
Eqs.~(\ref{psitilde1}) and (\ref{psitilde2}) are solved for the different 
values of $L$, running from $0$ to $L_{FSI}$. Since the excitation operator  
induces both isoscalar and isovector transitions, the hyperspherical harmonics 
(HH) entering the calculation are characterized by the quantum numbers 
$L=0$, $S=0$ and $T=0,1$. In the calculation involving $L$ up to $4$ the 
maximal value of the grand-angular quantum number $K_{max}$ is taken $7$ (odd multipoles)
or $8$ (even multipoles), the only exception being the $L=1$ multipole in the $T=0$ channel, 
for which $K_{max}=9$ has been used. For $L=5$ and $L=6$ $K_{max}$ is 
taken equal to $9$ and $10$, respectively. These values of the grand-angular 
quantum number provide the convergence of the various LIT's of Eq.~(\ref{mod}) 
with an uncertainty in the response function (\ref{FL}) of less than  $1\%$. 
In addition for $K_{max}=9$ and $10$ a selection of states has been performed 
with respect to the permutational symmetry types of the HH. Among the HH entering 
the expansion, those belonging to the irreducible representations [f]=[2] and 
[f]=[-]~\cite{ELO:1997A,FE:1981} of the four-particle permutation group 
{\bf S}$_4$ can be neglected in the calculation of the LIT for $K$ values 
higher than $7$ (odd multipoles) and $8$ (even multipoles).         

\begin{figure}
\resizebox*{12cm}{17cm}{\includegraphics{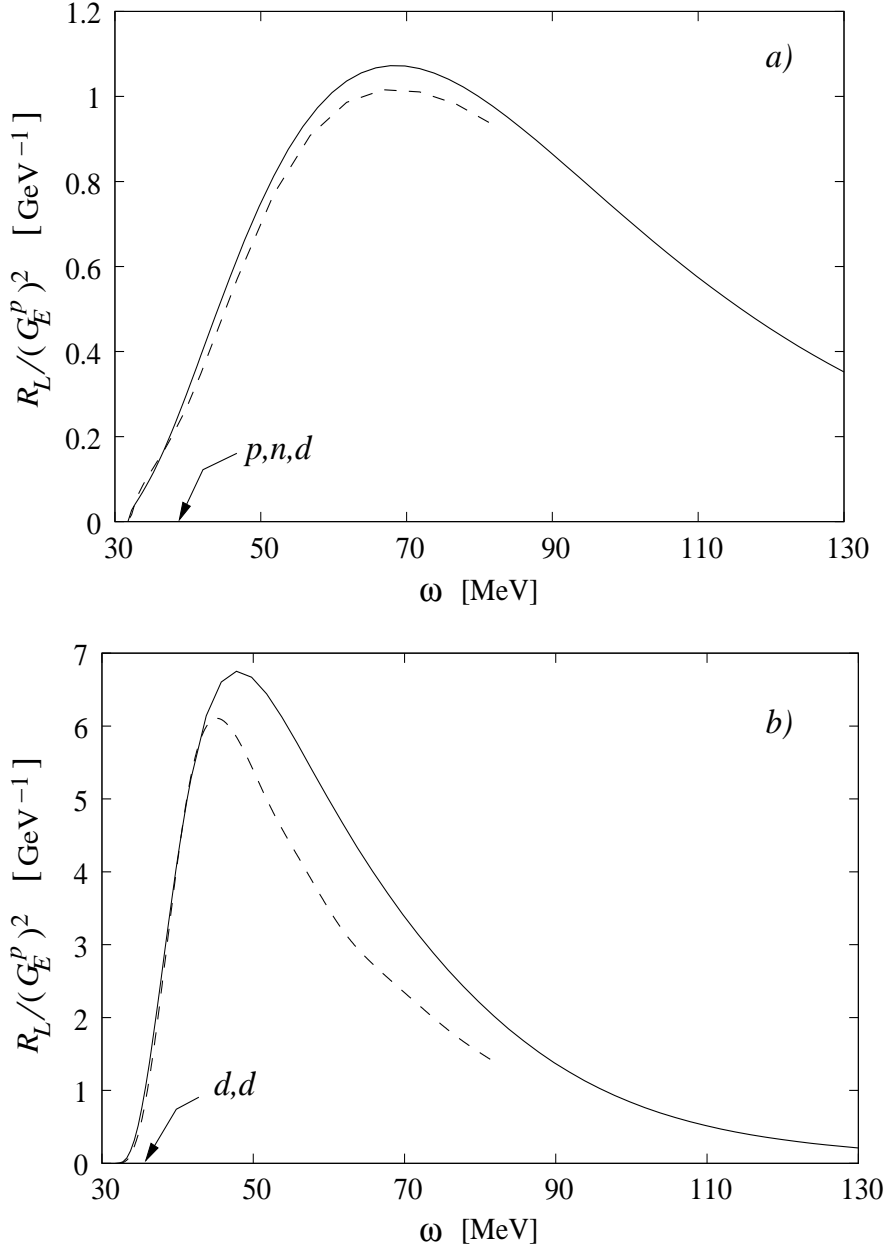}}
\caption{$T=1$, $L=0$ [panel {\it a)}] and $T=0$, $L=2$ [panel {\it b)}] components of 
the inclusive response (solid line) compared to those of $R^{p,t}_L+R^{n,h}_L$ 
[see Eq.~(\ref{RLpt})] (dashed line) for $q=300$ MeV/$c$. The arrows indicate the 
proton-neutron-deuteron (p,n,d,) and deuteron-deuteron (d,d) break--up thresholds.} 
\label{figure1}
\end{figure}

We start illustrating the contributions of the proton--triton channel and of 
the mirror channel due to the neutron--$^3$He break--up to the total inclusive 
response function. This comparison serves as a test of the correctness of the results.  
In fact below the threshold for the disintegration of $^4$He into proton, neutron 
and deuteron
for the isovector case and into two deuterons for the isocalar case,  the two results 
should coincide.
The neutron--$^3$He response $R_L^{n,h}(\omega,q)$ has been calculated along 
the same lines described above,
except that in the "channel state" $\phi_{f=a,b}^-(E_{f=a,b})$ of Eq.~(\ref{Psiminus}) 
the relative motion is given by a plane wave.
We choose to compare the sum of $R_L^{p,t}(\omega,q)$ and 
$R_L^{n,h}(\omega,q)$ for the multipoles L=0, T=1 and L=2, T=0 (two of the multipoles
which contribute most) with the total inclusive response calculated for the same 
multipoles.  In Fig.~\ref{figure1} this comparison is shown. Considering that  
the calculation of the total longitudinal response proceeds in a very different way, 
i.e. only by inversion of the norm of $\tilde \Psi_1$ \cite{ELO:1997A}, this 
comparison confirms the correctness of the calculation. Besides the degree of 
accuracy of the results one notices that for these multipoles the proton--triton 
and neutron--$^3$He channels dominate much beyond those thresholds. 

\subsection{Parallel kinematics}

%In the choice of the kinematics for 
Our study of $F_L(\omega,q,\theta_p)$ focuses first on the parallel kinematics of 
Ref.~\cite{DBB:1993}
where a Rosenbluth separation has been performed.  
In the columns 2-4 of Table~\ref{table1} we list the values of $q,\omega$ and modulus of 
missing momentum ${\bf p}_m={\bf q}-{\bf p}_p$ of the kinematics we have chosen to analyze
(labelled by Kin. N. in column 1). 
The values of the experimental energies
and momentum transfers  are illustrated in Fig.~2 as points in the 
$q - \omega$ plane and labelled with the corresponding numbers.
In the same figure their positions with respect to the
ridge $\omega=q^2/(2 m_p)$ are shown ($m_p$ is the proton mass).
The value of the final state intrinsic energy $E_{p,t}$, which is the input of the
calculation, has been obtained by calculating first the relative momentum ${\bf k}$
from the relation
\begin{equation}
{\bf k}=\mu\left(\frac{{\bf p}_p}{m_p} - \frac{({\bf q}-{\bf p}_p)}{m_t}\right)
\end{equation} 
and then using Eq.~(\ref{Ept}). 

In column 5 of Table~\ref{table1} the  PWIA results are listed. In this
approximation and in an independent particle model of $^4$He the PWIA result 
represents the probability that the proton
in the S-shell of $^4$He has momentum $p_m$. Therefore one has 
constant values for Kin. N. 1-3 and 4-8. The integral over all values of
${\bf p}_m$ gives the ``spectroscopic factor'' for that shell, which for
this potential turns out to be 0.88 \cite{ELO:1998} (this value can be compared with 0.84
obtained using a realistic potential like AV18 and 
Urbana IX~\cite{SPW:1986,Wir:1991,APW:1995}).

In Table~\ref{table1} the effects of 
antisymmetrization and of FSI are also shown as percentages of the 
PWIA values (the results denoted as FULL include both effects). In general one notices  
small effects of antisymmetrization 
for almost all cases as one would expect for kinematics with $p_m$ much smaller 
than $q$.
Nevertheless for the kinematics at lower energies these effects
can increase up to about 10\%. The role of FSI is much more important,
especially at low $q$. One notices that i) for the kinematics close to the  
$\omega=q^2/(2 m_p)$ ridge FSI effects  decrease for increasing  $q$;
ii) Kin. N. 4 and 9, which are more distant from the ridge $\omega=q^2/(2 m_p)$,  
present a rather high contribution of FSI; iii) at higher momenta and in the lower 
energy side of the ridge
%beyond 500 MeV/c, though for kinematics N.9 corresponding to
%a higher $p_m$ presents a rather high 
%contribution of FSI (about 50\%). 
FSI enhances the PWIA results. This effect goes in the opposite direction compared to
 previous estimates 
based either on optical potentials and orthogonalization 
procedure~\cite{SCH:1990}, or
on diagrammatic expansions~\cite{LAG:1989}. 

The observation iii) is consistent with previous ab initio calculations 
of the inclusive longitudinal response function in $^2$H 
~\cite{ALT:2005}, $^3$He ~\cite{GSWGNK:2005}) and $^4$He~\cite{ELO:1997A}. 
In the latter case one finds that the longitudinal responses at constant 
$q$ values, calculated with and without FSI, cross  at an $\omega$ value of 
approximately $q^2/(2 m_p)$. The fact that the crossing
happens just along that ridge is probably due to the different effects of 
the potential in the initial and in the final state with respect to the free one--body 
knock--out model, as explained in the following.
In the one body knock--out model the PWIA peak energy is $\omega_{peak}=q^2/(2 m_p)+\Delta$. 
The positive quantity $\Delta$ is the difference between 
the binding energies of $^4$He and $^3$H and can be considered as a "ground state effect" 
of the potential. One can argue that the additional effect of the potential in the final 
state would lead to $\omega_{peak}=q^2/(2 m_p)+\Delta-\bar V$, where $\bar V$ represents 
the mean interaction energy between the proton and the triton in the final state 
interaction zone ($\bar V$ will be attractive). Therefore the PWIA and FSI curves should 
intersect at an energy smaller than $q^2/(2 m_p)+\Delta$. To a good accuracy this 
value turns out to be just $q^2/(2 m_p)$. Of course such a comparison between inclusive 
and exclusive results is justified only in case of sufficiently low $p_m$ as it is the 
case for the kinematics listed in Table~\ref{table1}.

Similar PWIAS and FSI effects are also found for Kin. N. 6 and 9 in the two--body 
break--up results of $^3$He \cite{GKWGI:1995}.

%(the interesting region for "short range correlation" effects) 
It is a common belief that the kinematical regions at lower energy and higher momentum 
transfers 
are the privileged ones to investigate the ground state short range correlation
effects. Our results show that if one relies on approximate approaches to estimate the FSI 
effects one might underestimate considerably the momentum distributions at high $ p_m$ 
extracted from experiment in those kinematical regions.

As stated above, the aim of the present work is mainly to study relative effects of 
antisymmetrization and FSI, which are often treated approximately, via a complete 
solution of the quantum mechanical few-body problem. We have conducted this study 
using a semirealistic potential model. Nevertheless it is interesting to compare 
our results with experiment.
This  comparison is shown in Table~2.
Except for the case at the lowest $\omega$ and $q$ (Kin. N. 1)
where there is a good agreement, our results are almost systematically
higher than data. The difference ranges from about 30\% for the 
kinematics closer to the quasi elastic ridge 
to about 70 and even 100\% for the other ones. This comparison is better illustrated 
in Fig.~\ref{figure3}. One can see that, while FSI tends to bring theoretical results 
closer to data for the kinematics at lower momenta (Kin. N. 1-5), it affects in the 
opposite direction those at higher $q$ (Kin. N. 6-9), with the largest effect for  
Kin. N. 9 which corresponds to the highest $q$ and $p_m$-values. This is a delicate 
region where cross sections are small and potential dependence and relativistic effects
 neglected here might play a major role.

\subsection{Non parallel kinematics}

It is interesting to investigate the above effects also in
non parallel kinematics.
At fixed energy and momentum transfer one can access different $p_m$ varying $\theta_p$.
Therefore in PWIA the response reflects a proton momentum distribution. In the following 
we will show how  antisymmetrization and FSI can spoil this interpretation.  For the 
$(\omega,q)$ values of Table~\ref{table1} in Fig.~\ref{figure4}  results for 
non parallel kinematics are
shown as functions of $p_m$.
In the upper panel one can clearly see that the mere antisymmetrization effect
does not allow the interpretation of the response in terms of momentum distribution beyond
certain values of $p_m$, depending on the kinematics. These values are rather small (around 1 fm$^{-1}$)
for the kinematics at lower momentum transfer and can reach 2 fm$^{-1}$ for those
at higher $q$. This is of course discouraging for a study of the short
range correlations, which  contribute mainly to the higher tail of the momentum distribution.

Fig.~\ref{figure4} shows that antisymmetrization effects tend to fill the minimum of the response
in PWIA.
In order to illustrate the FSI effect, in the lower panel we have chosen Kin. N. 3 with  
a smaller and Kin. N. 9 with a larger $q$--value. As in parallel kinematics FSI tends 
to decrease the response in the former case and to enhance it in the latter. 
It is interesting to see that some minima reappear and some are filled when FSI 
is included. 

For a better understanding of the situation it is instructive to plot the 
matrix elements calculated from Eqs. (\ref{TfEf}), (\ref{TBorn})
and (\ref{TFSI1}). As an example we choose Kin. N. 3.
In Fig.~\ref{figure5}a our results for $T^{Born}_{p,t}$ are shown for PWIA and PWIAS. 
Moreover, in order to see the difference between an 
independent particle model and a correlated one, we also show the corresponding results obtained in 
an harmonic oscillator (h.o.) model. (The h.o. parameters have been fixed to the radii 
of $^4$He and $^3$H). Since the MTI--III potential has a rather strongly repulsive
core the comparison exhibits the effect of ground state short range correlations.
One readily sees that at low $p_m$ the MTI--III potential gives a 15\% quenching. 
The tail region is amplified in the inset. The results of the two models have similar behaviors
with increasing $p_m$, both in PWIA and PWIAS (see also inset of Fig.~\ref{figure5}a). 
However, while the h.o. PWIA matrix element remains 
always positive the corresponding one for MTI--III
crosses the zero axis, giving origin to the minimum 
visible in Fig.~\ref{figure4}. The minimum is then washed out by the antisymmetrization effect.

%Both h.o. and MTI--III curves in PWIA decrease 
%with increasing $p_m$. However while the h.o. result remains always positive the MTI--III curve
%crosses the zero axis giving origin to the minimum 
%visible in Fig.~\ref{nonparallel}. Also the effect of antysimmetrization is similar in the two cases
%making the result increase beyond $p_m=450 MeV/c$. However
%after  one also sees that, while the h.o. result 
%remains always positive beyond $p_m=450 MeV/c$, the effect of the MTI--III potential tends 
%to lower the h.o. result and crosses the zero at some point, giving origin to the minimum 
%visible in Fig.~\ref{nonparallel}. The effect of antisymmetrization is similar in the two 
%cases, tending to
%enhance the matrix element. The enhancement in the MTI--III case is such to wash out the 
%minimum.

In  Fig.~\ref{figure5}b the additional role of FSI is shown. In this case the total matrix 
element $T^{Born}_{p,t}+T^{FSI}_{p,t}$ is complex. Real and imaginary parts are 
shown and compared to the Born result with the MTI--III potential. In the inset the 
complicate interplay of the different contributions is illustrated. It is evident that FSI 
leads to a result  close to zero for a rather wide $p_m$ range, causing appearances and disappearances
of minima in the cross section. A more realistic interaction may change the present
picture in that kinematical region considerably. Nonetheless this model study points out that it
might be difficult to search for ground state correlation effects at high $p_m$ values
within a PWIA picture.

\section{Conclusions}

We have presented the results of an {\it ab initio} calculation of the 
$^4$He$(e,e^\prime p)^3$H longitudinal response obtained by means of the 
integral transform method with a Lorentz kernel. As NN interaction the MTI-III potential 
model is used. The aim has been to investigate
the limits of the PWIA approximation (factorization in terms of momentum distribution) 
due to the effects of antisymmentrization and FSI. We have analyzed the situation 
for the parallel kinematics investigated in the experiments of Ref.~\cite{DBB:1993} 
and for two non parallel kinematics.
Our model study has shown that the factorized approach (PWIA) might be a reasonable 
approximation for small missing momenta (below 0.5 fm$^{-1}$) and higher momentum transfers
(above 2 fm$^{-1}$). Unfortunately the situation for higher missing momenta becomes
much more involved. Both antisymmetrization effects and FSI play an important role.
In particular for non parallel kinematics their entanglement can give rise to drastic
deviations from the PWIA result. Furthermore, one may expect considerable 
sensitivity to nuclear dynamics here.
On the one hand this result can be considered discouraging in relation to the
possibility to "measure" directly  short range ground state correlations.
On the other hand it is possible that, due to the sensitivity of the response to
all effects, those kinematical regions are ideal
to study potential model dependence, including perhaps that due to thre--body forces.
However, FSI has to be treated in a proper way and realistic interactions
have to be used before definite conclusions can be drawn. The integral transform 
approach with a Lorentz kernel is a promising approach to pursue such studies.

\section*{Acknowledgment}
This work was supported by the grant COFIN03 of the Italian Ministery of University 
and Research. V.D.E. acknowledges support from the RFBR, grant 05-02-17541.

\newpage
%\bibliography{bib}

\newpage

%%%%%%%%%%%%%%%%%%%%%%%%%%%%%%%%%  TABLES %%%%%%%%%%%%%%%%%%%%%%%%%%%%%%%%%%%%%%%%%%%%%%%%%%%%%%%%%%%%%%%%%%%%%
\begin{table}[t]
\begin{center}
\begin{tabular}{cc c c c r@{.}l c ccr@{~}l c c c c c c c r @{.} l cc r @{~} cr@{.}l @{} ccc r @{~}r @{.}lc}
\hline
\hline
&&&&&\multicolumn{2}{c}{$ $}&&&&&&&
\multicolumn{8}{c}{PWIA}&&&\multicolumn{4}{c}{$ $}&&&&
\multicolumn{4}{c}{$ $}\\
Kin.&&&$q$&&\multicolumn{2}{c}{$\omega$}&&\multicolumn{4}{c}{$p_m$}&&
\multicolumn{8}{c}{$F_L/(G_E^{\,p})^2$}&&&
\multicolumn{4}{c}{$\Delta_{\rm PWIAS}$}&&&&
\multicolumn{4}{c}{$\Delta_{\rm FULL}$}\\
N.&&&$({\rm MeV}/c)$&&\multicolumn{2}{c}{(MeV)}&&\multicolumn{4}{c}{$({\rm MeV}/c)$}&&
\multicolumn{8}{c}{$[\left({\rm GeV}/c\right)^{-3}{\rm sr}^{-1}]$}&&&
\multicolumn{4}{c}{$(\%)$}&&&&
\multicolumn{4}{c}{$(\%)$}\\
\hline
$1$ &&& 299 && 57&78 &&&& $+$ & 30 &&&&&&&& $185$ & $2$ &&&$ $& $+$ & $9$&$3$  &&&& $-$ & $39$&$6$& $ $\\
                   	     				                  
$2$ &&& 380 && 83&13 &&&& $+$ & 30 &&&&&&&& $185$ & $2$ &&&$ $& $+$ & $1$&$2$   &&&& $-$ & $20$&$1$& $ $\\
                   	     				                  
$3$ &&& 421 && 98&19 &&&& $+$ & 30  &&&&&&&& $185$ & $2$ &&&$ $& $+$ & $0$&$0$   &&&& $-$ & $12$&$8$& $ $\\
                   	     				                  
$4$ &&& 299 && 98&70 &&&& $-$ & 90  &&&&&&&& $100$ & $0$ &&&$ $& $+$ & $4$&$5$   &&&& $-$ & $43$&$4$& $ $\\
                   	     				                  
$5$ &&& 380 && 65&06 &&&& $+$ & 90  &&&&&&&& $100$ & $0$ &&&$ $& $+$ & $3$&$9$   &&&& $-$ & $16$&$6$& $ $\\
                   	     				                  
$6$ &&& 544 && 126&6 &&&& $+$ & 90  &&&&&&&& $100$ & $0$ &&&$ $& $-$ & $1$&$1$   &&&& $+$ & $11$&$4$& $ $\\
                   	     				                  
$7$ &&& 572 && 137&82 &&&& $+$ & 90  &&&&&&&& $100$ & $0$ &&&$ $& $-$ & $1$&$9$   &&&& $+$ & $11$&$5$& $ $\\
                   	     				                  
$8$ &&& 650 && 175&67 &&&& $+$ & 90  &&&&&&&& $100$ & $0$ &&&$ $& $-$ & $1$&$7$   &&&& $+$ & $10$&$8$& $ $\\
                   	     				                  
$9$ &&& 680 && 146&48 &&&& $+$ & 190  &&&&&&&& $14$  & $63$ &&&$ $& $-$ & $4$&$4$   &&&& $+$ & $52$&$1$& $ $\\
\hline
\hline
\end{tabular}
\caption{The $^4$He$(e,e^\prime p)^3$H longitudinal response function of Eq.~(\ref{FL}) in PWIA 
approximation for the parallel kinematics of Fig.~2. The relative PWIAS and FULL 
effects, $\Delta_{\rm X}=\left({\rm X}-{\rm PWIA}\right)/{\rm PWIA}$, are also listed.
\label{table1}}
\end{center}
\end{table}

\begin{table}
\begin{center}
\begin{tabular}{ccccc r@{.}l c@{$\pm$} r@{.}l c@{$\pm$} r@{.}l cccc r@{.}l}
\hline\hline
Kin.&&&&&\multicolumn{14}{c}{$F_L$ $[\left({\rm GeV}/c\right)^{-3}{\rm sr}^{-1}]$}\\
No.&&&&&\multicolumn{8}{c}{Expt.}&&&&\multicolumn{3}{c}{FULL}\\
\hline
1&&&&& 59&0&& 2&0 && 2&2 &&&&& 66&4\\
2&&&&& 49&6&& 2&1 && 2&1 &&&&& 66&9\\
3&&&&& 46&2&& 2&5 && 2&2 &&&&& 62&7\\
4&&&&& 27&8&& 1&0 && 1&2 &&&&& 34&9\\
5&&&&& 28&4&& 1&2 && 1&3 &&&&& 37&2\\
6&&&&& 14&8&& 1&4 && 1&2 &&&&& 25&9\\
7&&&&& 16&0&& 1&5 && 1&3 &&&&& 23&0\\
8&&&&& 9&96&& 1&29&& 1&15&&&&& 16&3\\
9&&&&& 1&35&& 0&22&& 0&22&&&&& 2&73\\
\hline\hline
\label{table2}
\end{tabular}
\caption{The $^4$He$(e,e^\prime p)^3$H longitudinal response function. Theoretical results 
(FULL) compared to the experimental values of Ref.~\cite{DBB:1993}: statistical and 
systematic uncertainties are indicated ($\pm$stat. $\pm$syst.)}
\end{center}
\end{table}

%%%%%%%%%%%%%%%%%%%%%%%%%%%%%%%%%%%%% FIGURES  %%%%%%%%%%%%%%%%%%%%%%%%%%%%%%%%%%%%%%%%

\begin{figure}
\begin{center}
\resizebox*{12cm}{8cm}{\includegraphics{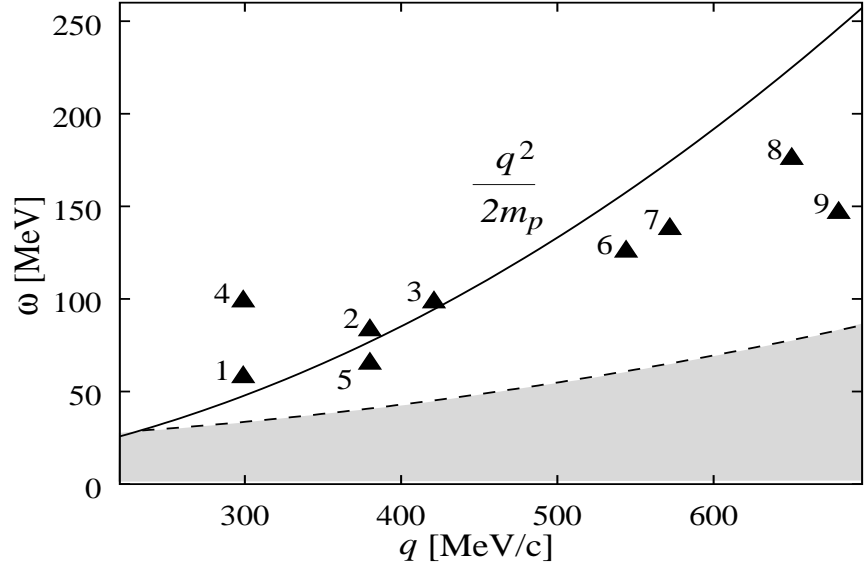}}
\caption{Position of the various kinematics of Table~\ref{table1} with respect to the  
$\omega=q^2/(2 m_p)$ ridge. The shaded area represents the region below the break--up 
threshold, where the cross section is zero.}
\end{center}
\label{figure2}
\end{figure}

\begin{figure}
\resizebox*{12cm}{8cm}{\includegraphics{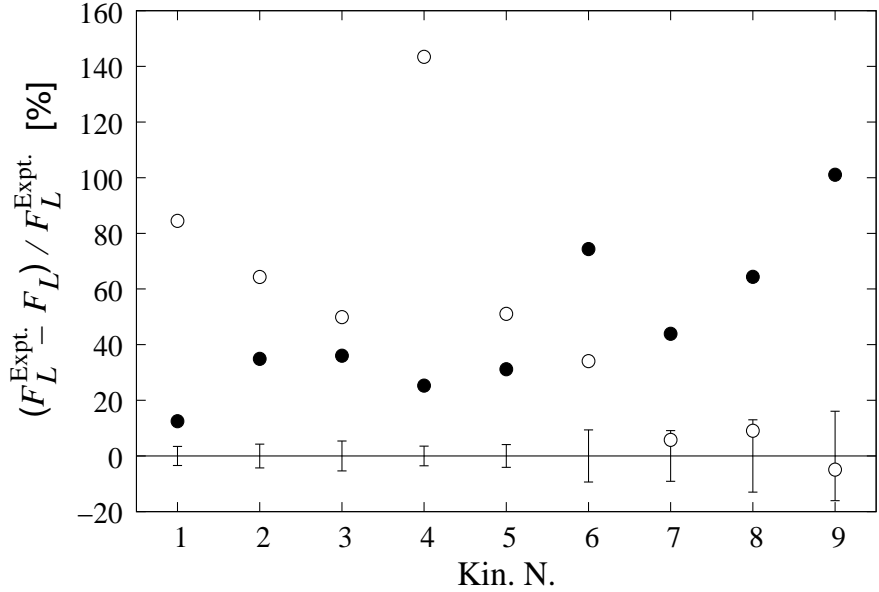}}
\caption{Percentage deviation from the experimental values of Ref.~\cite{DBB:1993}: 
PWIA (open circles), FULL results (full circles).}
\label{figure3}
\end{figure}

\begin{figure}
\resizebox*{12cm}{17cm}{\includegraphics{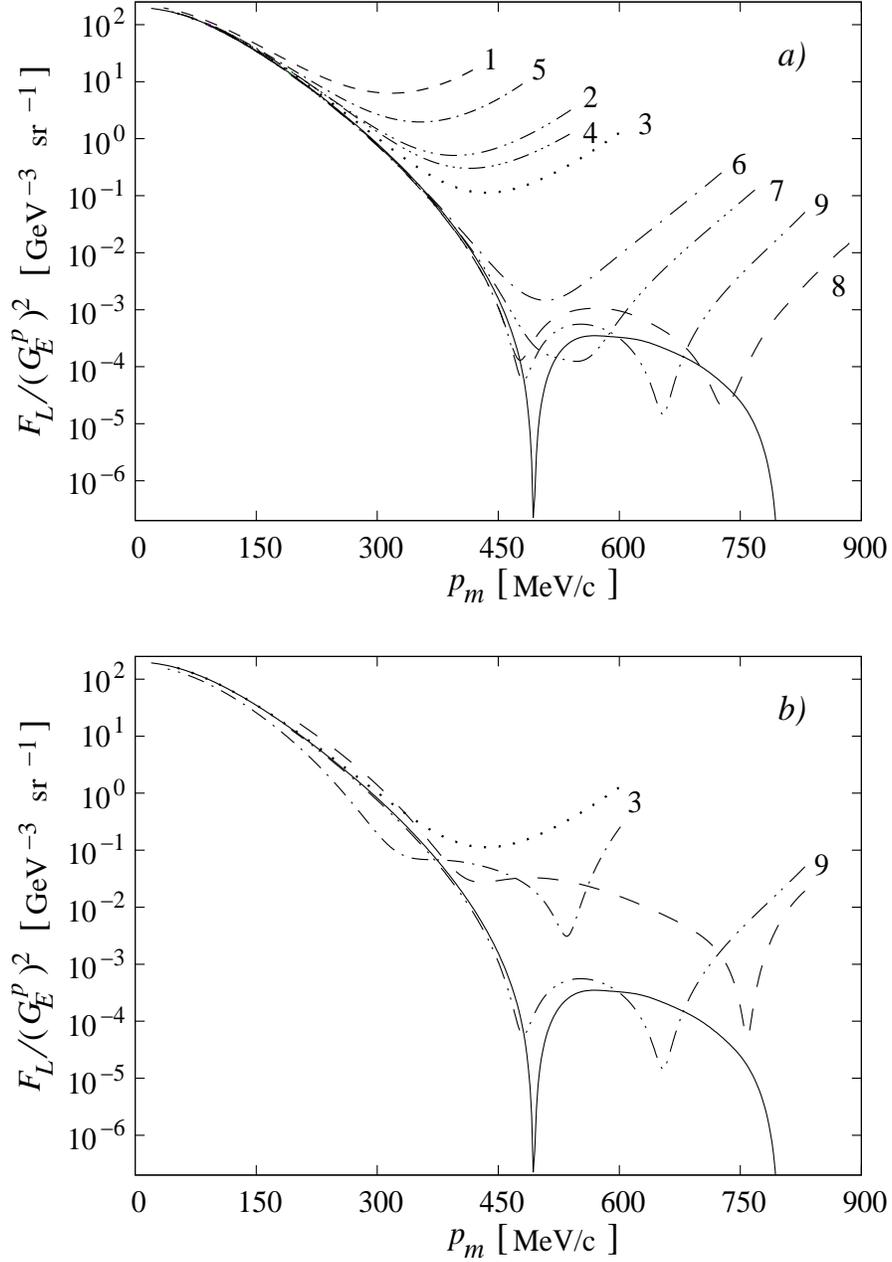}}
\caption{The $^4$He$(e,e^\prime p)^3$H longitudinal response function of Eq.~(\ref{FL}) 
as function of $p_m$: {\it a)} PWIAS for the different $(\omega,q)$ values of
Table~{\ref{table1}} labelled with the corresponding numbers; {\it b)} PWIAS (dotted and dot--dot--dashed 
line) and FULL (dashed and dot--dashed line) results for the $(\omega,q)$ values of 
Kin. N. 3 and 9 of Table~\ref{table1}. The solid line 
represents the PWIA.} 
\label{figure4}
\end{figure}

\begin{figure}
\resizebox*{12cm}{17cm}{\includegraphics{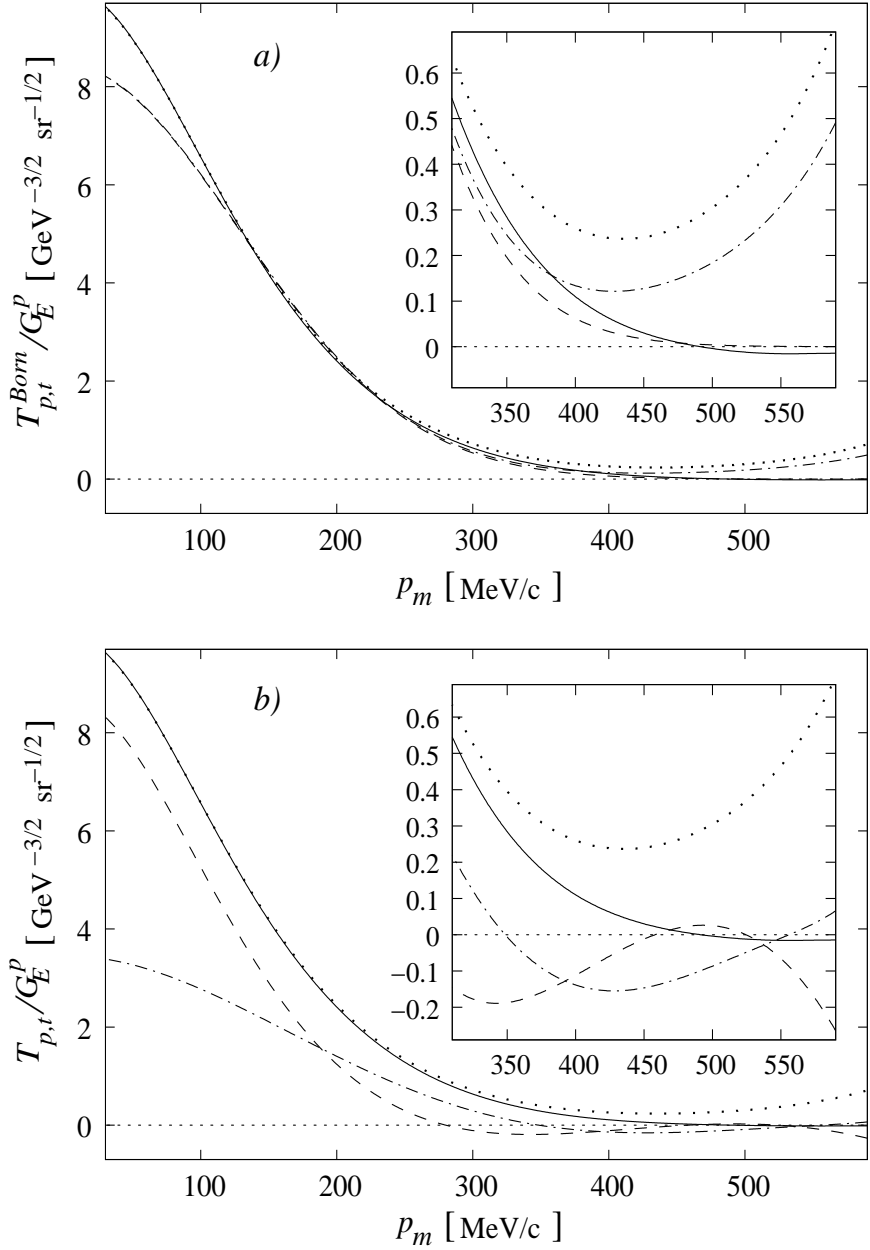}}
\caption{Matrix element $T(E)$ of Eq.(\ref{Tpt}) in non parallel kinematics, for $q$ and $\omega$  
of Kin. N. 3, as function of $p_m$. a): h.o. model: PWIA (dashed line), 
PWIAS (dotted line); MTI--III potential model: PWIA (full line),  MTI--III,
PWIAS (dot--dashed line). b) MTI--III potential model: real (dashed line) and imaginary 
(dot-dashed line) parts of the FULL 
matrix element. PWIA and PWIAS results  as in a).}
\label{figure5}
\end{figure}

\end{document}